\documentclass[aps,prb,twocolumn,letterpaper,superscriptaddress,showpacs]{revtex4-1}
\usepackage{graphicx}
\usepackage{CJK}
\usepackage{amsmath}
\usepackage{color}
\usepackage{dcolumn}
\usepackage{bm}
\usepackage{hyperref}
\usepackage{txfonts}

\newcommand{\cfno}{(Co$_{0.97}$$^{57}$Fe$_{0.03}$)$_{4}$Nb$_{2}$O$_{9}$}
\newcommand{\cno}{Co$_{4}$Nb$_{2}$O$_{9}$}
\newcommand{\fto}{Fe$_{4}$Ta$_{2}$O$_{9}$}

\begin{document}
\begin{CJK*}{UTF8}{bsmi}
\title{
Modulated non-collinear magnetic structure of \cfno\ as revealed \\ by M\"ossbauer spectroscopy
}
\author{Bo Zhang}
\affiliation{Key Lab for Magnetism and Magnetic Materials of the Ministry of Education, Lanzhou University, Lanzhou 730000, China}
\author{Qifeng Kuang}
\affiliation{Shenyang National Lab for Materials Science, Institute of Metal Research, Chinese Academy of Sciences, and School of Materials Science and Engineering, University of Science and Technology of China, 72 Wenhua Road, Shenyang, 110016, China}
\author{Hua Pang}
\author{Fashen Li}
\author{Liyun Tang}
\affiliation{Key Lab for Magnetism and Magnetic Materials of the Ministry of Education, Lanzhou University, Lanzhou 730000, China}
\author{Da Li}
\affiliation{Shenyang National Lab for Materials Science, Institute of Metal Research, Chinese Academy of Sciences, and School of Materials Science and Engineering, University of Science and Technology of China, 72 Wenhua Road, Shenyang, 110016, China}
\author{Zhiwei Li}\email{zweili@lzu.edu.cn}
\affiliation{Key Lab for Magnetism and Magnetic Materials of the Ministry of Education, Lanzhou University, Lanzhou 730000, China}

\date{\today}

\begin{abstract}
In this work, we present detailed $^{57}$Fe M\"ossbauer spectroscopy investigations of \cfno\ compound to study its possible magnetic structure. We have shown that the previously reported magnetic structures can not satisfactorily describe our low temperature M\"ossbauer spectra. Therefore, in combination with theoretical calculations, we have proposed a modulated helicoidal magnetic structure that can be used to simulate the whole series of our low temperature M\"ossbauer spectra.
Our results suggest that the combination of previously reported different magnetic structures are only approximations of the average magnetic structure from our modulated helicoidal model.
We anticipate that the proposed modulated non-collinear magnetic structure might shed light on the understanding of the complex magnetoelectric effects observed in this system.
\end{abstract}


\maketitle
\end{CJK*}

\section{Introduction}
Recently, the corundum-type compounds, M$_4$A$_2$O$_9$ (M = Mn, Co, Fe, A = Nb, Ta) \cite{prb.103.014422,prb.98.024410,prm.2.091401r,prb.93.075117,prb.102.174443,prb.103.L140401}, have drawn great interests due to their rich physics such as magnetoelectric (ME) effect \cite{prb.103.014422,prb.98.024410}, magnetic anisotropy \cite{prb.103.014422,prb.102.174443} and magnetism induced spontaneous electric polarization in \fto\ \cite{prb.98.024410,prm.2.091401r}.  Among these compounds, \cno , which is the most studied one, has been found to show large ME effect below the N\'{e}el temperature of $T_N\sim27\,K$ \cite{apl.99.132906,sr.4.3860}. Cross coupling between the electric field induced magnetization and magnetic field controlled electric polarization have been experimentally observed on powder and single crystalline samples \cite{sr.4.3860}.
The microscopic mechanisms underlining these interesting properties remain to be settled. It is believed that the detailed magnetic structures play very important roles in understanding these interesting properties \cite{prb.103.L140401,prb.103.014422,prb.94.094427,prb.97.020404r,prb.97.085154,jpsj.88.094704}.

The crystal structure of \cno\ is rather simple and have been investigated both by polycrystalline and single crystal X-ray and neutron diffraction methods \cite{jpcs.21.234, prb.93.075117,prb.97.085154,prb.102.174443}.
The magnetic structure was first reported by Bertaut et al. \cite{jpcs.21.234} to be antiferromagnetically coupled ferromagnetic Co$^{2+}$ chains with their magnetic moments along the crystal $c$-axis. It has been shown later by Khanh et al. \cite{prb.93.075117} that this magnetic structure is incompatible with the experimentally measured ME effect on single crystal samples. A different magnetic structure with all the magnetic moments nearly parallel to the [1\={1}0] direction has been proposed based on single-crystal neutron diffraction. Within this model, the magnetic moments exhibit a small canting towards the $c$-axis but the projected moments on the $ab$-plane are all parallel with each other.
Later on, both powder and single crystal neutron diffraction data have revealed another distinct non-collinear magnetic structure without any moment canting to the $c$-axis \cite{prb.97.085154,prb.102.174443}. Moreover, different spin-flop behaviors have been reported in \cno\ and magnetic structures associated with these magnetic anomalies have been suggested to explain the observed ME effect \cite{apl.99.132906,sr.4.3860,prb.93.075117}.

Clearly, one has to know the precise magnetic structure to understand the observed interesting ME effects. To reconcile the above mentioned diverse magnetic structures \cite{jpcs.21.234,prb.93.075117,prb.97.085154,prb.102.174443}, we present the results of a local probe of the magnetic structure of \cfno\ compound with M\"ossbauer spectroscopy. Electric field gradient (EFG) tensor has been calculated by density functional theory and served as the coordinate system to determine the directions of the magnetic moments. Tentative fittings of our M\"ossbauer spectra with previously published magnetic structures have all failed. Therefore, we proposed a more complex helicoidally modulated magnetic structure to satisfactorily describe the whole series of our M\"ossbauer spectra. The proposed non-collinear magnetic structure may be promising in the understanding of the complex ME observed in related systems.

\section{Experimental}
Polycrystalline samples of \cfno\ and \cno\ were prepared by using the conventional solid state reaction technique \cite{apl.99.132906}. To prevent reaction of the sample with oxygen, both mixing and reaction procedures were made inside an argon glove box with the oxygen level controlled below 0.1\,ppm. A total of $\sim$1\,gram Co/$^{57}$Fe, Co$_3$O$_4$, and Nb$_2$O$_5$, in a proportion to meet the correct oxygen content, were used and the reaction temperature was fixed at 1100\,$^o$C. Several reactions with thorough intermediate grindings were needed to improve homogeneity of the doped $^{57}$Fe and to reduce foreign phases such as CoNb$_2$O$_6$.

Phase purity was checked by room temperature X-ray powder diffraction (XRPD) and the refinements were done by using the FullProf suite \cite{fullprof}. Magnetic properties were measured using a dc superconducting quantum interference device (SQUID) magnetometer (Quantum Design). M\"ossbauer measurements were performed in transmission geometry with a conventional spectrometer working in constant acceleration mode. A 50\,mCi $\gamma$-ray source of $^{57}$Co embedded in Rh matrix and vibrating at room temperature was used. The drive velocity was calibrated using $\alpha$-Fe foil for high velocity measurements and sodium nitroprusside (SNP) for low velocity measurements. The isomer shift quoted in this work are relative to that of the $\alpha$-Fe at room temperature.

The computational work was carried out within the \textsc{ELK} code \cite{elk}, which is based on the full potential linearized augmented plane waves (FP-LAPW) method. The Perdew-Wang/Ceperley-Alder local spin density approximation (LSDA) exchange-correlation functional \cite{prb.45.13244} was used. LSDA+U calculation was done in the fully localized limit (FLL) and by means of the Yukawa potential method \cite{prb.52.R5467} with a screening length of $\lambda=3.0$. Slater integrals are calculated according to $\lambda$ and the resulting Coulomb interaction parameters are $U=3.0$\,eV and $J=0.88$\,eV which are similar to previously used values \cite{prb.94.094427}. The muffin-tin radii $R_{MT}$ were set to 2.33\,a.u., 2.07\,a.u., and 1.40\,a.u. for Co, Nb, and O atoms, respectively. The plane-wave cutoff was set to $R_{MT}\times |\textbf{G}+k|_{max} = 7.0$ and the maximum \textbf{G}-vector for the potential and density was set to $|\textbf{G}|_{max}=12.0$. $\textbf{k}$-point meshes of $4\times4\times2$ a total of 32 k-points were used due to very time consuming calculations when spin orbital coupling (SOC) were included. Experimental lattice parameters of \cno\ at 50\,K ($a = b = 5.180$\,{\AA} and $c = 14.163$\,{\AA}) were used for our calculation \cite{prb.102.174443}.

\section{Results and discussion}
\begin{figure}[ht]
\includegraphics[width=0.9\columnwidth,clip=true]{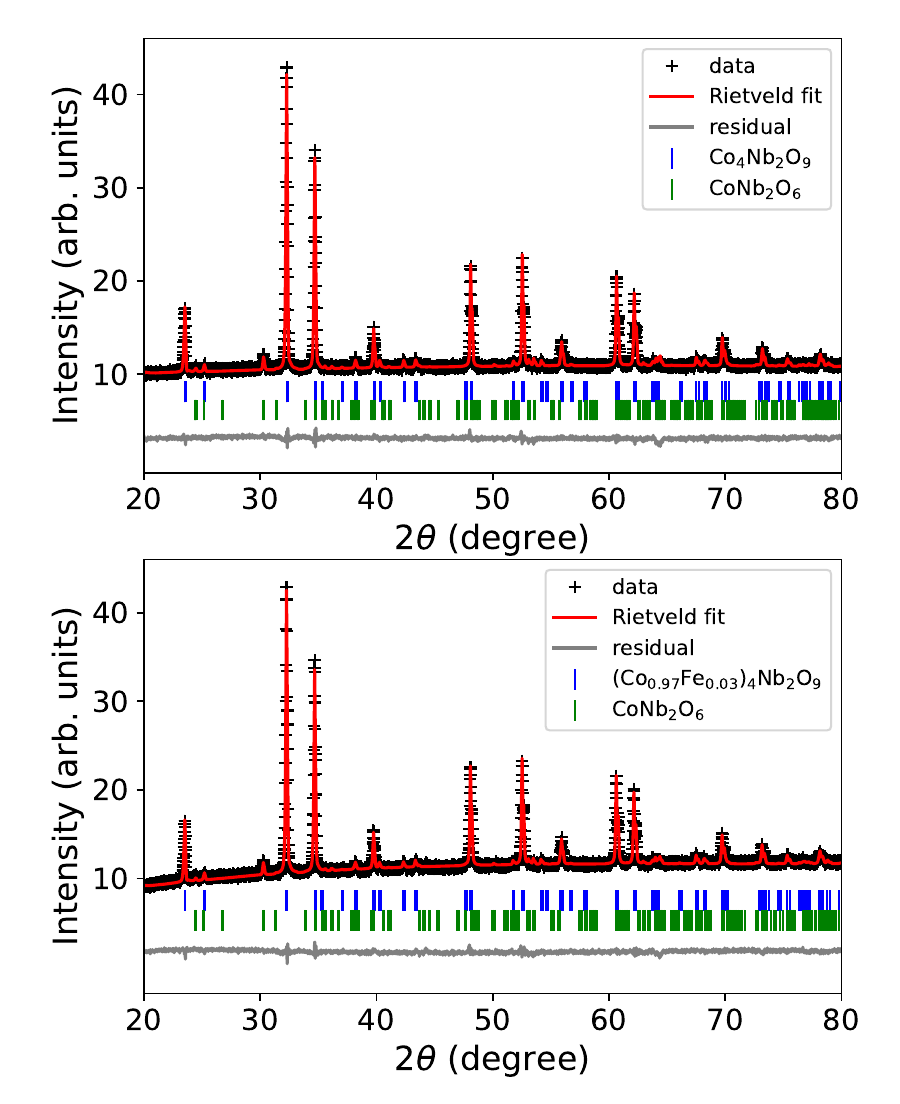}
\caption{\label{fig:XRD}
(color online)
Rietveld refinements of the room temperature X-ray powder diffraction data of (a) \cno\ and (b) \cfno .
}\end{figure}

In Fig. \ref{fig:XRD}, the Rietveld refinements of the room temperature XRPD data for both the pure \cno\ compound for comparison and the $^{57}$Fe doped \cfno\ are shown. P\={3}c1 space group characteristic of the desired 429-phase \cite{jpcs.21.234,apl.99.132906} with a small amount of CoNb$_2$O$_6$ impurity 126-phase \cite{ci.47.14041}, which is usually found in the title compound, were used to refine our XRPD data.
The amount of the impurity 126-phase were refined to be about 1.8\,wt\% and 1.4\,wt\% for \cno\ and \cfno, respectively.
In agreement with earlier reports \cite{jmcc.9.14236,apl.99.132906}, the determined lattice parameters are $a = 5.1702(1)$ and $c = 14.1306(4)$ for \cfno, which are a little larger than the values of $a = 5.1680(1)$ and $c = 14.1268(4)$ for \cno\ due to bigger ionic radius of Fe$^{2+}$ than Co$^{2+}$, indicating the successful doping of the $^{57}$Fe atom into the lattice of the parent 429-phase.

\begin{figure}
\includegraphics[width=1.0\columnwidth,clip=true]{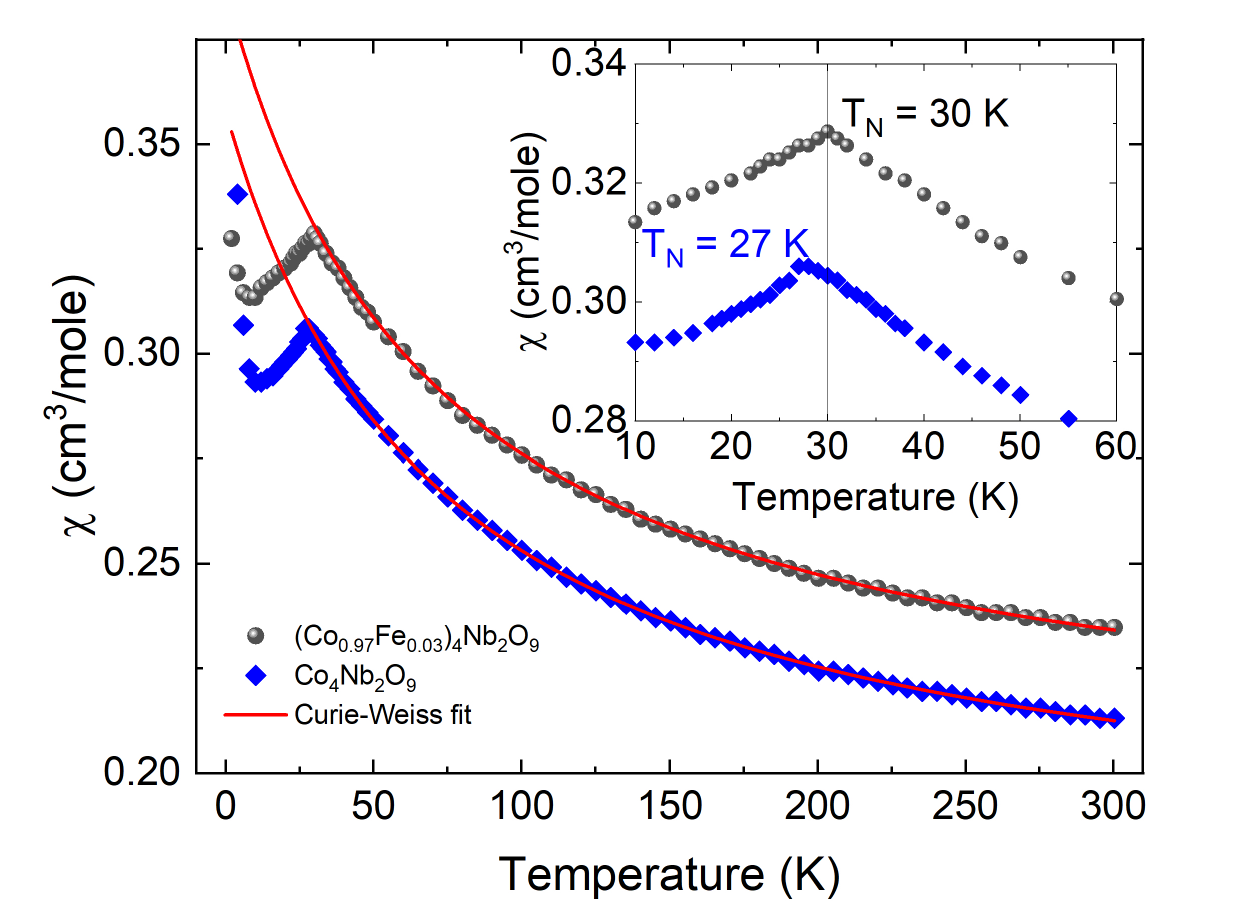}
\caption{\label{fig:MT}
(color online)
Temperature dependence of the magnetic susceptibility ($\chi$) measured in zero-field-cooling mode together with Curie-Weiss fitting (red solid curve) for \cno\ and \cfno. Inset: enlargement of the $\sim$30\,K region showing the N\'{e}el transition temperature, $T_N$.
}\end{figure}

The temperature dependence of the magnetic susceptibility, $\chi (T)$, measured in zero-field-cooling mode shown in Fig. \ref{fig:MT} were fitted with the Curie-Weiss law $\chi (T) = \chi_0 + C/(T + \theta)$ in the paramagnetic regime. The fitted Weiss temperatures are $-72\,K$ and $-71\,K$ and the effective magnetic moments ($\mu_{eff}$) are $5.10\,\mu_B$ and $5.15\,\mu_B$ for \cno\ and \cfno, respectively. The obtained $\mu_{eff}$ values are close to previously reported values and indicate the high spin state of the Co$^{2+}$ ions with $S = 3/2$ \cite{prb.93.075117}. However, these values are significantly larger than the spin only effective magnetic moments of $\sim$3.87\,$\mu_B$ for $S = 3/2$, indicating unquenched contribution from the angular momentum to the effective magnetic moments via strong spin orbital coupling \cite{prb.93.075117, prb.94.094427}. As shown in the inset of Fig. \ref{fig:MT}, the N\'{e}el transition temperatures were found to be $T_N\sim27$\,K and $T_N\sim30$\,K for \cno\ and \cfno, respectively. The slightly enhanced $T_N$ suggests that the relatively small doping of $\sim3\%$ $^{57}$Fe is not totally effect-less to the parent compound \cite{jmcc.9.14236} but we expect that the impact should be small with such a small doping level for insulators. The upturn at low temperatures can be ascribed to a weak amount of paramagnetic impurities usually observed in polycrystalline samples that can not be seen by our XRPD and M\"ossbauer spectroscopy.

\begin{figure}
\includegraphics[width=1.0\columnwidth,clip=true]{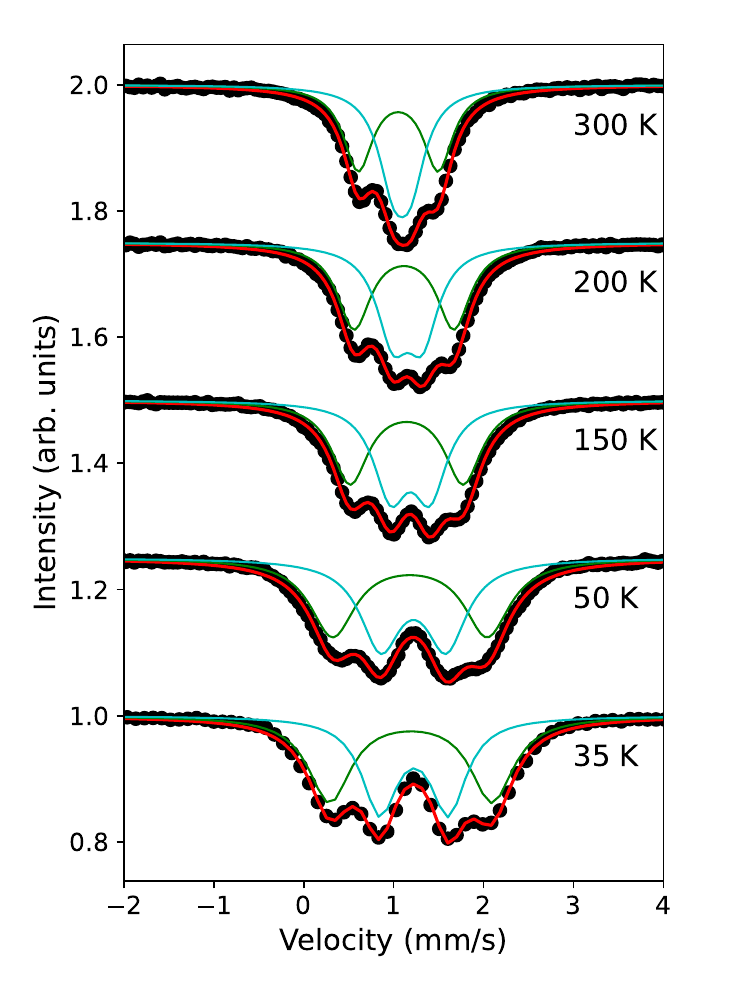}
\caption{\label{fig:Hmoss}
(color online)
$^{57}$Fe M\"ossbauer spectra (black dots) of \cfno\ measured at indicated temperatures above $T_N$. Solid lines are fits to the experimental data with two doublets as described in the text.
}\end{figure}

The $^{57}$Fe M\"ossbauer spectra of \cfno\ taken in the paramagnetic temperature range, $T>T_N$, shown in Fig. \ref{fig:Hmoss} were fitted with two doublets corresponding to the two Co/$^{57}$Fe sites in the crystal structure \cite{jpcs.21.234}, namely the relatively flat Co/$^{57}$FeO$_6$ layer, Co/$^{57}$Fe(1)/site 1, and the highly buckled Co/$^{57}$FeO$_6$ layer, Co/$^{57}$Fe(2)/site 2.
The fitted values of the isomer shift (IS) are $\delta_{1}(300\,K)=1.092(1)$\,mm/s and $\delta_{2}(300\,K)=1.050(1)$\,mm/s, respectively. These values are typical for high-spin ions Fe$^{2+}$ ($d^6, S = 2$) located in FeO$_6$ octahedrons \cite{PGmbook2011}. The values of the quadrupole splitting (QS) are $\Delta_{1}(300\,K)=0.192(4)$\,mm/s and $\Delta_{2}(300\,K)=0.889(2)$\,mm/s, respectively.
To identify the $^{57}$Fe doping site, we made calculation of the QS for the non-magnetic phase.
The EFG tensor was calculated by
\begin{equation}
V_{ij}^{\alpha} \equiv \frac{\partial^2V^{'}_C(\textbf{r})}{\partial \textbf{r}_i\partial \textbf{r}_j}|_{\textbf{r}=\textbf{r}_{\alpha}}
\end{equation}
where $V^{'}_C$ is the Coulomb potential with the $l=m=0$ component removed in each muffin-tin.
The calculated $V_{zz}(1) = -1.03\times 10^{21}$\,V/m$^2$ and $V_{zz}(2) = -3.60\times 10^{21}$\,V/m$^2$ with $\eta(1)=\eta(2)=0$.
The corresponding QS (with $\Delta=eQV_{zz}/2$) values are $\Delta_{c1}=-0.171$\,mm/s and $\Delta_{c2}=-0.599$\,mm/s for the flat Co/$^{57}$Fe(1)/site 1 and the buckled Co/$^{57}$Fe(2)/site 2, respectively. These values agree reasonably well with our experimental results.

With decreasing temperature, we observed obvious increasing of the separation between the two peaks of both doublets, corresponding to larger QS values of $\Delta_{1}(35\,K)=0.763(3)$\,mm/s and $\Delta_{2}(35\,K)=1.807(4)$\,mm/s.
These high QS values indicate that the $^{57}$Fe nuclei are located in crystal sites with stronger EFG, which reflects the asphericity of the charge-density distribution near the probing nucleus.
We also noticed that the fitting quality decreases with decreasing temperature with only two simple doublets, which may be caused by several reasons. For example, 1) short range magnetic correlations that appear well above $T_N$; 2) cluster relaxation effects due to inhomogeneous doping of the $^{57}$Fe etc., which will be investigated in the future with more measurements in this temperature range.

\begin{figure*}
\includegraphics[width=2\columnwidth,clip=true]{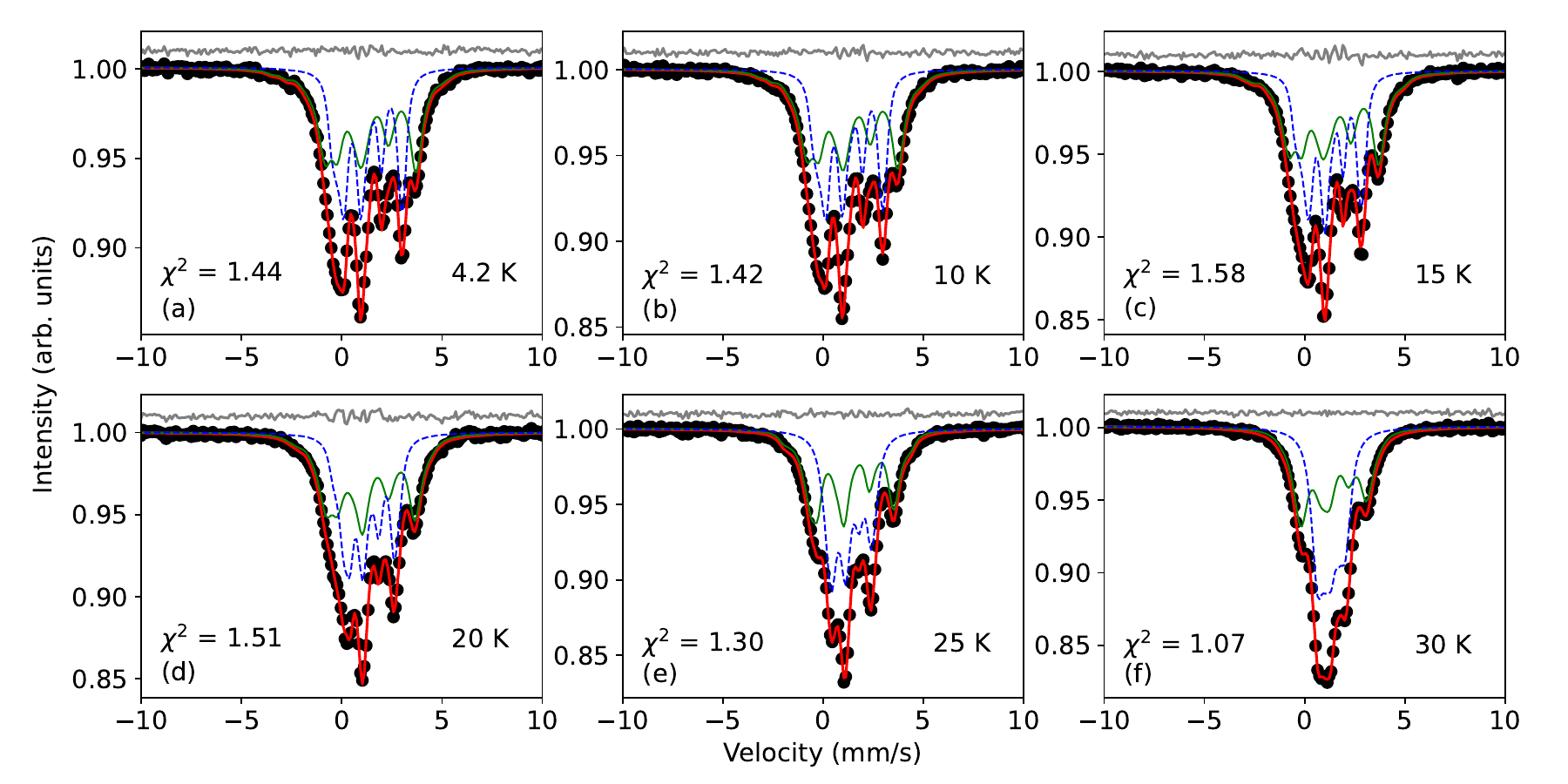}
\caption{\label{fig:Lmoss}
(color online)
(a)-(f) $^{57}$Fe M\"ossbauer spectra (black dots) of \cfno\ measured at indicated temperatures below $T_N$. Solid red lines are simulation of the experimental spectra as described in the text. Dashed blue lines and solid green lines are sub-spectra corresponding to $^{57}$Fe sitting at site1 and site2, respectively. Differences between the experimental data and the simulated curves are shown as solid gray lines above the spectra. The resulting $\chi^2$ are smaller than $\sim$1.58 for all these spectra as indicated in the figure.
}\end{figure*}

As shown in Fig. \ref{fig:Lmoss}, below $T_N \approx 30$\,K, complex Zeeman splittings appear in our M\"ossbauer spectra. To fit these spectra, considering that the QSs are large, we used the full Hamiltonian of hyperfine interactions in the coordinate system of the principal axes ($|V_{zz}|\geq|V_{yy}|\geq|V_{xx}|$) of the EFG tensor \cite{PGmbook2011}
\begin{eqnarray}
\label{eqHamilton}
\mathcal{\hat{H}}_{QM}&& = \mathcal{\hat{H}}_Q + \mathcal{\hat{H}}_M \\ \nonumber
&& = \frac{eQV_{zz}}{4I(2I-1)}[3\hat{I}^2_z - \hat{I}^2 + \eta(\hat{I}_x^2 - \hat{I}_y^2)] \\ \nonumber
&& - g\mu_NB[(\hat{I}_x\cos\phi + \hat{I}_y\sin\phi)\sin\theta + \hat{I}_z\cos\theta],
\end{eqnarray}
where $\hat{I}$ and $\hat{I}_x, \hat{I}_y, \hat{I}_z$ refer to the nuclear spin operator and operators of the nuclear spin projections onto the principal axes and $Q$ denotes the quadrupole moment of the nucleus. $\phi$ and $\theta$ are the azimuthal and polar angles of the hyperfine magnetic field in the EFG coordinate system, respectively. $\eta=(V_{yy}-V_{xx})/V_{zz}$ represents the asymmetry parameter of the EFG at the nucleus.

It is important to note that, the eigenvalues of such Hamiltonian depend on a number of parameters, such as ($eQV_{zz}$, $B$, $\eta$, $\theta$, and $\phi$), and some of them are correlated. Thus, these parameters cannot be determined independently from M\"ossbauer spectra, but only in certain combinations.
Moreover, the analysis of the M\"ossbauer spectra can only determine the direction of the magnetic moments in the coordinate system of the principal axes of the EFG tensor. It is necessary to know the relative direction of the EFG coordinate system with respect to the crystal lattice coordinate before we can determine the magnetic structure by fitting of our M\"ossbauer spectra.

Therefore, before any detailed analysis of our low temperature M\"ossbauer spectra, we made theoretical analysis based on first-principle electronic structure calculations to study the EFG tensor and other hyperfine interactions.
First, we did calculations without considering SOC. To establish the collinear in-plane antiferromagnetic structure as a starting point, an internal magnetic field was applied within the muffin-tin of Co atoms in the [$\pm1, 0, 0$] direction which was used to break the symmetries and was reduced to almost zero in the end by setting a reducing factor of 0.95 for each self-consistent field calculation circle. The converged magnetic moments are spin only in this case and amounts to $\mu_{S1}=2.63$\,$\mu_B$ and $\mu_{S2}=2.59$\,$\mu_B$ for the two Co atoms. These values are actually close to the reported values from single-crystal neutron diffraction \cite{prb.102.174443}.
The calculated $V_{zz}(1) = -8.15\times 10^{21}$\,V/m$^2$ and $V_{zz}(2) = -8.29\times 10^{21}$\,V/m$^2$ with $\eta(1)=\eta(2)=0$. Then, the QS was calculated by $\Delta = eQV_{zz}/2(1+\eta^2/3)^{1/2}$ to be $\Delta_{1}=-1.36$\,mm/s and $\Delta_{2}=-1.38$\,mm/s. These values are inconsistent with the values determined at 35\,K from above both in magnitude and relative ratio of the two sites which may be an indication of redistribution of the electron charges at the probing sites that happens when going through the antiferromagnetic transition. The magnetic hyperfine field for each atom was calculated according to ref. \cite{prb.35.3271} implemented within the \textsc{ELK} code. The fermi contact field were calculated to be $B_{F}(1)=32.0$\,T and $B_{F}(2)=30.6$\,T. These values are much too large when compared with the magnetic splittings of our M\"ossbauer spectrum at 4\,K as shown in Fig. \ref{fig:Lmoss} (a). When the spin dipole part was considered, these values get even larger to be $B_{FD}(1)=36.2$\,T and $B_{FD}(2)=33.6$\,T. Since the hyperfine field is mainly contributed by three parts as $B_{hf} = B_F + B_D + B_L$, we have to consider the orbital part $B_L$ in the present case. In fact, for high-spin Fe$^{2+}$, $B_L$ can be as large as $\sim 20$\,T and opposite to $B_F$ \cite{PGmbook2011}.

Therefore, we made calculations with SOC included. In this case, an internal magnetic field was applied within the muffin-tin of Co in the [$\pm1, 0, \delta_l = 0.057$] direction to break the symmetries, which reduces to nearly zero as in our earlier calculations without SOC.
Free rotation of the local moment towards $c$-axis and within the $ab$-plane were both allowed due to Dzyaloshinskii-Moriya like interactions as a result of SOC. Due to this moment rotation, our calculation converges extremely slow, indicating that the SOC plays an important role in determining the ground state of \cno. However, after several thousands of iterations, we found that the total energy changes of the system are very small, $\sim 0.1$\,meV. Most importantly, we found that the changes of the EFG tensor is negligibly small, e.g. the change of $V_{zz}$ is smaller than $\sim$1\,$\%$ between the last 2000 iterations.
From our calculation, the main component ($V_{zz}$) of the EFG tensor is roughly directed along the $c$-axis of the crystal structure, with a maximum deviation angle of $\sim$3.3\,$^o$ for all the 8 Co atoms in the unit cell.
Therefore, we fixed the direction of the magnetic moments to the direction of the experimental easy axes of the magnetic moments determined by a tentative fit of the 4.2\,K M\"ossbauer spectrum. In this case, our calculation was easily converged to a level of $\sim 0.27$\,$\mu eV$.

From our DFT results, it is reasonable to assume that the direction of $V_{zz}$ is along the $c$-axis in the fitting procedure of our M\"ossbauer spectra. According to previously reported magnetic structures \cite{prb.93.075117, prb.97.085154, prb.102.174443}, we tried to fit our M\"ossbauer spectra measured below $T_N$ with three different models. Model I): canting of the magnetic moments out of the $ab$-plane and free rotation between the two sites are both allowed. Model II): canting of the magnetic moments out of the $ab$-plane is allowed but with a single projected direction in the $ab$-plane for the two sites \cite{prb.93.075117}. Model III): the magnetic moments are confined within the $ab$-plane but free rotation between the two sites are allowed \cite{prb.97.085154, prb.102.174443}.
According to the above three models, two sextets were fitted to the experimental data by considering different directions of the hyperfine magnetic field within the coordinate system of the principal axes of the EFG tensor. This is in accordance with the commensurate propagation vector \textit{k}=0 found in the above three models. However, as shown in Fig. S1 - S3 in the supplemental material, all the above mentioned three models exhibit obvious discrepancies from the experimental data at low temperatures. Actually, only the spectrum taken at 30\,K can be reasonably described with a magnetic structure similar to the one proposed by Khanh et el. \cite{prb.93.075117}. With decreasing temperature, the quality of the fits get worse gradually as indicated also by the $\chi^2$ values shown in Fig. S1 - S3.
Since our investigated sample was actually doped with $^{57}$Fe, there might be local randomness that introduced by possible inhomogeneous doping effects. This may give a distribution of the measured hyperfine magnetic field. Indeed, we also tried to fit our M\"ossbauer spectra using the above models by including hyperfine field distribution effects \cite{nim.B58.85}. But, one can see from Fig. S4 that there is no improvement when compared with the simple two sextets model. Actually, the temperature evolution of the M\"ossbauer spectra also rules out the possibility that local randomness to be the reason why our fitting got worse and worse with decreasing temperatures.
These results suggest that the magnetic structure of \cfno\ or even \cno\ changes from a nearly collinear simple structure \cite{prb.93.075117} at higher temperature to a more complex possibly non-collinear one at lower temperatures.

Therefore, to describe the low temperature spectra satisfactorily, we used a more sophisticated model by taking into account the features associated with an incommensurate helicoidal structure as were used to described the complex M\"ossbauer spectra of BiFeO$_3$ \cite{nm.12.641}, 3$R$-AgFeO$_2$ \cite{jpcm.29.275803}, Fe$_3$PO$_7$ \cite{prb.97.104415}, FeP \cite{jac.675.277} and many others with modulated magnetic structures. In this case, the experimental spectrum is approximated as a superposition of two sets of Zeeman patterns corresponding to the two crystal sites. $\phi$ and $\theta$ angles in equation \ref{eqHamilton} for each Zeeman pattern can be expressed by the rotation angle $\omega$ within the magnetic moment rotation plane, which varies continuously in the $0\leq \omega_i \leq 2\pi$ interval for an incommensurate helicoidal structure. The angle, $\Theta_n$, between the normal direction of the rotation plane and the $c$-axis/$V_{zz}$ was fitted as a free parameter. To take into account of possible anharmonicity (bunching) of the spatial distribution of the magnetic moments of Co$^{2+}$/$^{57}$Fe$^{2+}$, a Jacobian elliptic function was used \cite{jpcm.29.275803,jac.675.277}
\begin{equation}
\label{eqJacobian}
cos \omega (z) = sn([\pm 4 K(m)/\lambda]z, m)
\end{equation}
where $K(m)$ is the complete elliptic integral of the first kind, $\lambda$ the helicoidal period, and $m$ is the anharmonicity parameter related to the distortion of the helicoidal structure.

To take account the local field anisotropy that distorts from circular helicoid, we consider the anisotropic hyperfine coupling tensor \textbf{A}. We assume that \textbf{A} is diagonal with respect to the principal axes of the EFG tensor and the hyperfine field can be written as \cite{jpcm.29.275803}
\begin{eqnarray}
\label{aiBhf}
B_{hf} && = \textbf{A}\cdot \mathbf{\mu} = A_{xx}\mu_{x}\hat{x} + A_{yy}\mu_{y}\hat{y} + A_{zz}\mu_{z}\hat{z} \\ \nonumber
&& = A_{\perp}(\mu_x\hat{x} + \mu_{y}\hat{y}) + A_{\parallel}\mu_z\hat{z}
\end{eqnarray}
where $\mu_{x,y,z}$ are projections of the $^{57}$Fe spin moment on the principal axes of the EFG tensor, and $A_{\perp}\equiv A_{xx}=A_{yy}$, $A_{\parallel}\equiv A_{zz}$ were used since the direction of $V_{zz}$ coincides with the crystal $c$-axis.
The hyperfine field component projected onto the field rotation plane can be written as $B_{\omega_i}=\mu (A^2_{\parallel}cos^2\omega_i + A^2_{\perp} sin^2\omega_i)^{1/2}$. However, in first order, only the component $\textbf{A}\cdot \mathbf{\mu}$ parallel to the spin direction will affect the magnitude of the hyperfine field. Then we can rewrite the expression used in our fitting procedure as $B_{\omega_i}(\parallel \mu) = B_{\parallel}cos^2\omega_i + B_{\perp} sin^2 \omega_i$ \cite{jpcm.29.275803}. 
In our model, one of the anisotropy direction was assumed to be perpendicular to the plane formed by the easy magnetization direction and the $c$-axis/$V_{zz}$.
The easy direction of the magnetic moments for the two sites are described by four angles, namely $\Theta_{ea}(1)$/$\Theta_{ea}(2)$ and $\Phi_{ea}(1)$/$\Phi_{ea}(2)$. During our fitting procedure, we found that using two $\Theta_n(1)$/$\Theta_n(2)$ angles for the two rotation planes does not improve much of the fitting quality than using only one for both sites. Thus, in the final fits, one single $\Theta_n$ for the two rotation planes was used in the fitting model.
The asymmetry parameter, $\eta$, were also fixed to the calculated values of $\eta(1)=\eta(2)=0.15$ for both sites.
Moreover, due to the small value of $\eta$, we can not accurately determine the value of $\Phi_{ea}$, e.g. the fitted standard deviation is very large. Therefore, we have fixed $\Phi_{ea}$ to the 90$^o$ for both sites.

\begin{figure}
\includegraphics[width=0.8\columnwidth,clip=true]{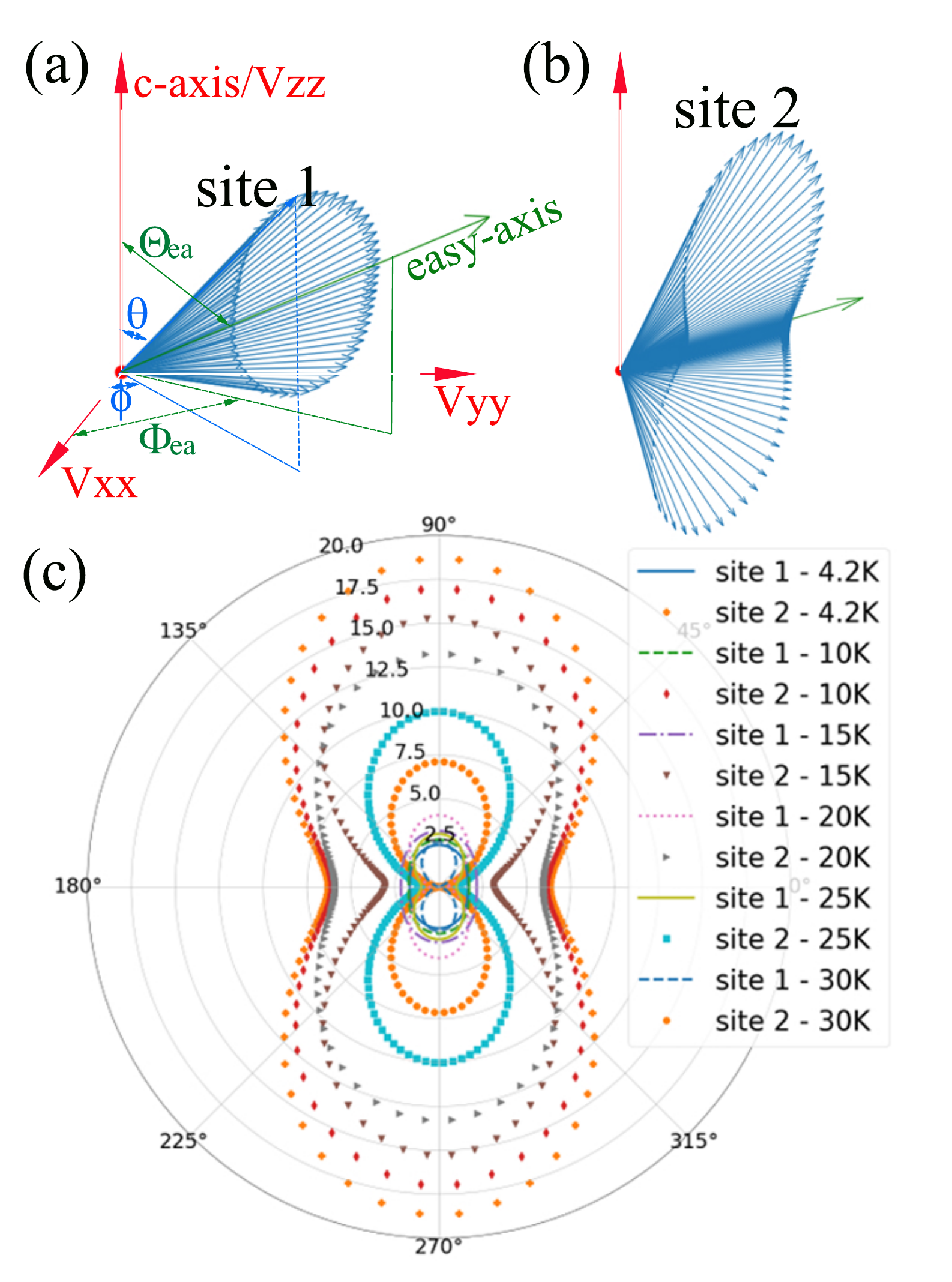}
\caption{\label{fig:helicoidal}
(color online)
(a) and (b) the corresponding helicoidal contour for Co/$^{57}$Fe(1) (site 1) and Co/$^{57}$Fe(2) (site 2) at 4.2\,K, respectively, determined from fitting of the spectrum shown in Fig. \ref{fig:Lmoss} (a). $\phi$ and $\theta$ are the angles for the magnetic moment within the EFG coordinate system defined in equation \ref{eqHamilton}. $\Theta_{ea}$ and $\Phi_{ea}$ are the angles of the easy magnetic direction within the EFG coordinate system used in the fitting procedure. Panel (c) shows the projections of the anisotropic hyperfine field on the moment rotation plane at indicated temperatures below $T_N$ (direction $0^o - 180^o$ is parallel to the $ab$-plane of the crystal structure).
}\end{figure}

One can see from Fig. \ref{fig:Lmoss}\ that using the above model allowed us to satisfactorily describe the entire series of our M\"ossbauer spectra below $T_N$. The fitting quality were found to be much better than previous models, that is, the resulting $\chi^2$ were reduced to be smaller than $\sim$1.58 for all these spectra. At 4.2\,K, the determined EFGs corresponding to the two sites are $V_{zz}(1) = -6.9(2)$\,V/m$^2$ and $V_{zz}(2) = -9.6(4)$\,V/m$^2$, which are close with the calculated values ($V_{zz}(1) = -7.89$\,V/m$^2$ and $V_{zz}(2) = -8.17$\,V/m$^2$ from our theoretical calculation with SOC.
Moreover, from our DFT calculation, the non-zero $\eta$ values were obtained only after SOC was included, indicating that the strong SOC has a considerable effect on the asymmetry distribution of the charges around the probing nucleus which deserves further theoretical studies.
The fitted directions for the easy magnetization are $\Theta_{ea}(1) = 74(1)$\,$^o$, $\Theta_{ea}(2) = 69(1)$\,$^o$ in the coordinate system of EFG. The normal direction of the rotation plane was found to be $\Theta_n = 105(1)$\,$^o$, resulting in angles of 31\,$^o$ and 36\,$^o$ relative to the corresponding easy magnetization directions of the two sites.
The fitted anharmonicity parameter $m$ at 4.2\,K are close to $\sim0$ and $\sim$\,1 for site 1 and site 2, respectively. With increasing temperature, $m$ quickly gets close to $\sim$\,1 for both sites. These values suggest that the modulated component of the magnetic moments are nearly within the $ab$-plane at higher temperatures close to $T_N$. This explains why the 30\,K spectrum can be well fitted with only two subspectra.
The corresponding helicoidal contour for the two sites at 4.2\,K, projected on the same atoms, are shown in Fig. \ref{fig:helicoidal} (a) and (b). Clearly, at ground state, the helicoidal structure is more isotropic for Co/$^{57}$Fe(1) but more anisotropic for Co/$^{57}$Fe(2). With increasing temperature, the magnetic structure eventually becomes almost collinear near the transition temperature, $\sim30$\,K.
The spatial anisotropy of the determined hyperfine magnetic field, $\textbf{B}_{hf} \propto \textbf{A}\bullet \mathbf{\mu_{eff}}$, is largely due to the angular dependency of the hyperfine coupling tensor \textbf{A} and/or anisotropy $\mathbf{\mu_{eff}}$ as were discussed in other systems \cite{prb.97.104415, jpcm.29.275803}.

\begin{figure}
\includegraphics[width=1.0\columnwidth,clip=true]{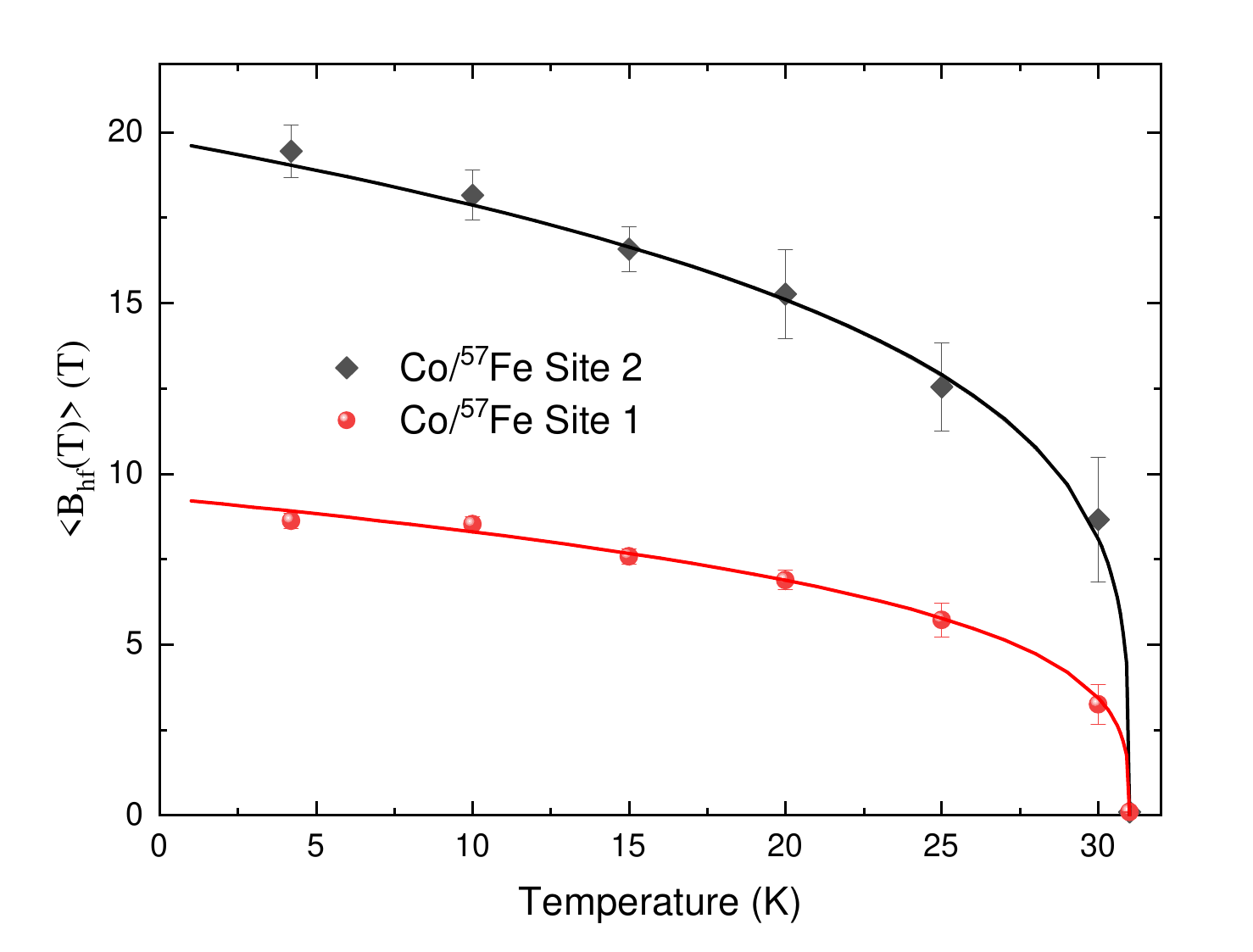}
\caption{\label{fig:BhfT}
(color online)
Temperature dependence of the average hyperfine magnetic field $\langle B_{hf}(T)\rangle$ for Co/$^{57}$Fe(1) and Co/$^{57}$Fe(2) sites of \cfno. Solid lines are power-law, $\langle B_{hf}(T)\rangle$ = B$_{hf}$(0)(1 - $T/T_N$)$^{\beta}$, fits to the experimental data in the temperature range between 15\,K and 30\,K.
}\end{figure}

The determined average hyperfine field, $\langle B_{hf}\rangle$, from their distribution $p(B_{hf}) \propto (|\partial B_{hf}(\omega)/\partial \omega|)^{-1}$ \cite{jac.675.277} were shown in Fig. \ref{fig:BhfT} as a function of temperature. Solid lines are power-law, $\langle B_{hf}(T)\rangle$ = B$_0$(1 - $T/T_N$)$^{\beta}$, fits to the experimental data in the temperature range between 15\,K and 30\,K. The fits lead to $B_{0}(1) = 9.3(3)$\,T, $\beta (1) = 0.29(3)$ and $B_{0}(2) = 20(2)$\,T, $\beta(2) = 0.26(6)$ for Co/$^{57}$Fe(1) and Co/$^{57}$Fe(2) sites, respectively. The values of the critical exponent are between the theoretical value of $\sim0.23$ expected for two-dimensional XY system and $\sim0.345$ for three dimensional XY system \cite{jpcm.20.275233, prb.21.3976}.
The fitted values of $B_{0}$ are much smaller than the fermi contact field, as mentioned above, but are much larger than the values from our calculation with SOC included ($B_{FDL}(1) = 3.4$\,T and $B_{FDL}(2) = 7.5$\,T), reflecting the delicate balance between the two contributions.


\begin{figure}
\includegraphics[width=0.85\columnwidth,clip=true]{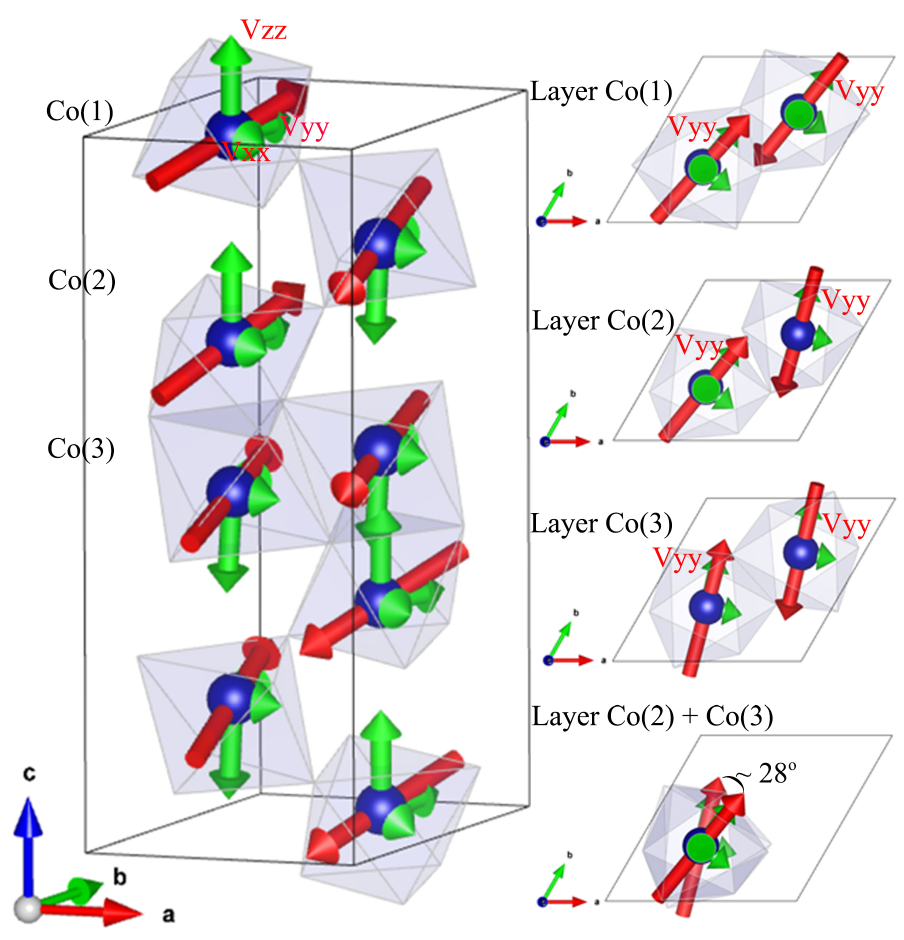}
\caption{\label{fig:magstr}
(color online)
Left panel shows the crystal structure together with (green arrows) calculated directions of the principal axes of the EFG tensor ($|V_{zz}|\geq |V_{yy}| \geq |V_{xx}|$), and (red arrows) easy axes of the magnetic moments determined from fitting of the 4.2\,K M\"ossbauer spectra shown in Fig. \ref{fig:Lmoss} (a). Right panel shows the top view of the corresponding layers that labeled on the left panel plotted using the VESTA package \cite{vesta}.. The directions of $V_{yy}$ are also labeled. The calculated rotation angle of $V_{yy}$ or the direction of the easy magnetization is $\sim$28\,$^o$ as labeled in the bottom of the right panel.
}\end{figure}

Finally, from the above discussed experimental and calculation results, we should try to understand the magnetic structure of the title compound. As discussed above, our fitting of the magnetic structure to the M\"ossbauer spectra were done in the coordinate system of the principal axes of the EFG tensor. Therefore, we show our calculated directions of the principal axes of the EFG tensor in the left panel of Fig. \ref{fig:magstr} as green arrows. The directions of the easy axes are shown together as red arrows. The top view of corresponding layers as labeled in the left panel are shown in the right panel. We can see that there is an inter-layer rotation between neighboring layers for the Co(1) site and an intra-layer rotation for the Co(2) site. The calculated angle amounts to $\sim$28$^o$ as shown in the bottom of the right panel. Clearly, the observed modulated helicoidal magnetic structure arises as a compromise of competing interactions, such as antiferromagnetic $J_1$ between Co(1)-Co(1), ferromagnetic $J_2$ between Co(1)-Co(2), and single ion anisotropy etc., similar to previously reported systems \cite{nm.12.641, prb.97.104415}. If we take the easy axis as the average direction of the magnetic moments as shown in Fig. \ref{fig:magstr}, one immediately finds out that it is a combination of previously reported magnetic structures by different groups with neutron diffraction techniques \cite{prb.93.075117, prb.97.085154, prb.102.174443}. The canting angles out of the $ab$-plane amounts to 16\,$^o$ for site 1 and 21\,$^o$ for site 2 , determined from our M\"ossbauer data, compares well with the magnetic structure reported by Khanh et al. \cite{prb.93.075117}. On the other hand, the in-plane rotation angle calculated in this work $\sim$28$^o$, which can not be determined by the current M\"ossbauer data, is much larger than the values reported by Ding et al. \cite{prb.102.174443}.
We also want to emphasize that our proposed modulated helicoidal magnetic structure is in contrast with previous models observed by neutron diffractions \cite{prb.93.075117, prb.97.085154, prb.102.174443}. This is not surprising to us since the reported magnetic structures are different from group to group even they were all deduced by neutron diffraction techniques. Moreover, these simple magnetic structures can not fully explain the observed complex ME effects \cite{apl.99.132906,sr.4.3860,prb.93.075117}. There are also accumulating evidence that the magnetic structures are much more complex than previously observed with neutron diffractions in similar systems of Fe$_4$Nb$_2$O$_9$ \cite{prb.103.L140401} and Co$_4$Ta$_2$O$_9$ \cite{prb.102.214404}. Especially, the observed spontaneous electric polarization in the absence of magnetic field in Fe$_4$(Nb,Ta)$_2$O$_9$ \cite{prm.2.091401r,prb.98.024410,ci.47.9055} systems are still lacking of an explanation.
Due to the complexity of the actual ground state magnetic structure and the similarity of the average magnetic structure to the reported models, it could be difficult to be refined by neutron diffraction techniques due to domain and powder averaging effects \cite{prb.92.134419}.
Thus, our results call for more precise measurements with specially designed neutron diffraction experiments to reveal the true magnetic ground states of this system.

\section{Summary}
Concluding, we have carried out $^{57}$Fe M\"ossbauer measurements on polycrystalline samples of \cfno\ to study its possible magnetic structures. Our results show that this compound exhibits very large electric field gradient (EFG), therefore, the principal axes of the EFG tensor can be used as a good coordinate system to solve its magnetic structure by fitting of the measured M\"ossbauer spectra. The directions of the principal axes ($|V_{zz}|\geq |V_{yy}| \geq |V_{xx}|$) were calculated by density functional theory with spin orbital coupling effect. From these theoretical calculations, we have proposed a modulated helicoidal magnetic structure to simulate the low temperature M\"ossbauer spectra since all other previously reported magnetic structures failed to describe these M\"ossbauer spectra. The average magnetic structure, derived from the fitted easy axis direction of the magnetic moments, can be used to reconcile previously reported conflicting magnetic structures. Our proposed non-collinear magnetic structure may be useful in the explanation of the complex magnetoelectric effects observed in this system.

\section{Acknowledgement}
Project supported by the National Natural Science Foundation of China (Grant Nos. 11704167, 51971221).
The authors are grateful to the support provided by the Supercomputing Center of Lanzhou University.

\bibliography{cfno}

\end{document}